\documentclass[12pt]{article}
\usepackage{amssymb}
\usepackage{amsfonts}
\usepackage{amsmath}
\usepackage[mathscr]{eucal}
\usepackage{amssymb}
\usepackage{amsthm}
\usepackage{mathrsfs}
\usepackage{bbold}
\usepackage{bm}
\usepackage{graphicx}
\usepackage{caption}
\usepackage[english]{babel}
\usepackage[T2A]{fontenc}
\usepackage[utf8]{inputenc}
\usepackage{cite}

\newcommand{\be}{\begin{equation}}
\newcommand{\ee}{\end{equation}}
\newcommand{\bea}{\begin{eqnarray}}
\newcommand{\eea}{\end{eqnarray}}

\newcommand{\lb}{\label}
\newcommand{\p}[1]{(\ref{#1})}
\def\theequation{\arabic{section}.\arabic{equation}}

\usepackage{MnSymbol}

\newcounter{rown}

\begin{document}
\begin{titlepage}
\vspace*{0.5cm}

\begin{center}
{\LARGE\bf BRST construction for infinite spin field on $AdS_4$}

\vspace{1.5cm}

{\large\bf I.L.\,Buchbinder$^{1,2,3}$\!\!,\,\,
S.A.\,Fedoruk$^1$\!,\,\,  A.P.\,Isaev$^{1,4}$\!,\,\,  V.A.\,Krykhtin$^{2}$}

\vspace{1.5cm}

\ $^1${\it Bogoliubov Laboratory of Theoretical Physics,
Joint Institute for Nuclear Research, \\
141980 Dubna, Moscow Region, Russia}, \\
{\tt buchbinder@theor.jinr.ru, fedoruk@theor.jinr.ru,
isaevap@theor.jinr.ru}

\vskip 0.5cm

\ $^2${\it Department of Theoretical Physics,
Tomsk State Pedagogical University, \\
634041 Tomsk, Russia}, \\
{\tt joseph@tspu.edu.ru, krykhtin@tspu.edu.ru}

\vskip 0.5cm

\ $^3${\it National Research Tomsk State  University,}\\{\em Lenin
Av.\ 36, 634050 Tomsk, Russia}

\vskip 0.5cm

\ $^4${\it Faculty of Physics, Lomonosov Moscow State University,
119991 Moscow, Russia}

\end{center}

\vspace{2cm}

\nopagebreak

\begin{abstract}
\noindent We generalize the first class constraints that describe the infinite spin irreducible $4D$ Poincar\'{e} group representation in flat space to new first class constraints in $AdS_4$ space. The constraints are realized as operators acting in Fock space spanned by the creation and annihilation operators which are two-component spinors. As a result, we obtain a new closed gauge algebra on $AdS_4$ with the known flat space limit. Using this gauge algebra, we construct the BRST charge and derive the Lagrangian and gauge transformations for free bosonic infinite spin field theory in $AdS_4$ space.
\end{abstract}

\vspace{2cm}


\phantom{Keywords: }

\newpage

\end{titlepage}

\setcounter{footnote}{0}
\setcounter{equation}{0}

\section{Introduction}

Higher spin field theory attracts much attention during long time
although many fundamental problems in this area are still open. To
the present time in the framework of general higher spin theory,
several more narrow directions have been formed closely related both
with fundamental problems and with each other (see e.g. the reviews
\cite{revVas,revBCIV,reviews3,rev_Bekaert,DidSk,reviewsV,Snowmass,Ponomarev}
and the reference therein). One of such directions is a problem of
Lagrangian description for infinite spin field.

As it is known, the massless irreducible representations of the
Poincar\'{e} group in flat space are divided into helicity
representations (representations with define helicities) and
infinite (continuous) spin representation \cite{Wig,BargWig}.
Lagrangian description of helicity representations is well
developed, at least for free fields and for cubic field interactions
(see e.g. the recent paper \cite{BuchResh} and the references
therein). As for the representation of infinite spin, interest in
its study has arisen relatively recently and various aspects of its
Lagrangian description are currently being examined
\cite{BekSk,BekMur,SchTor,Met,Met1,Met2,Zin1,Na,Zin,AG,BFIR,BKT,BIF,BKSZ,BFIK,BIFP,BFIK-24}.
It should be noted however, that most of the research in this area
concerns fields in flat space. In the present paper we will develop
a general approach to Lagrangian formulation for free bosonic
infinite spin field in the four dimensional AdS space. Our
approach is based on formulation with use of
canonical BRST-charge (to be more precise the BRST-charge in
BFV construction \cite{BFV} (see also, the reviews \cite{BFH} and the book \cite{HT}).
Further, as is usually customary in higher spin theory, we use the term BRST-charge, meaning the
BFV-construction).

The infinite spin fields in AdS space were discussed in the review
\cite{BekSk}, where a generic description of such fields was given,
and in the refs. \cite{Met,Met2,Zin}. In the ref. \cite{Met}, the
Lagrangian infinite spin formulation was postulated by
adding some new terms both to sum of the Fronsdal Lagrangians on AdS and to
corresponding gauge transformations to mix the fields with
different helicities. This result was later confirmed in \cite{Met2} within light-cone approach. In ref. \cite{Zin}, the frame-like
Lagrangian description for infinite spin field in AdS space was realized on
the base of massive finite spin field theory on AdS space in the
limit when mass tends to zero and spin to infinity, but
their product remains fixed. Also, it was pointed out in the refs.
\cite{BekSk,Met,Met2,Zin} that the consistent Lagrangian
formulation for infinite spin field exists not in arbitrary curved
space-time but only in AdS space.

Recently, we have constructed  \cite{BFIK-24} a new model of relativistic world line particle
in four dimensional constant curvature space based on
generalization of the gauge algebra for infinite spin particle in
flat space to curved space-time. In the current paper starting from such a particle model
we develop a general approach to Lagrangian description for infinite spin field in $4D$ AdS space.
Unlike the approaches in refs. \cite{Met,Met2,Zin}, we work in
the framework of BRST-BFV construction, where a central role is
played by first class constraint algebra. Such an algebra has never been presented in the literature
before. We emphasize that this algebra cannot be obtained by simply
covariantizing the corresponding algebra of the theory in flat
space, since the consistent algebra must include special non-minimal
terms proportional to curvature.

It is worth pointing out that the notion of an infinite spin field
in AdS space in the context under consideration needs clarification.
The point is that the infinite spin irreducible representation
exists only in Minkowski space. When we talk about a dynamic
formulation of an infinite spin field in $4D$ AdS space, we actually
mean a theory with a closed gauge algebra that coincides in the flat
limit with the corresponding gauge algebra for the theory of an
infinite spin field in $4D$ Minkowski space\footnote{One can say
that the field on AdS space is referred to as
\textit{continuous-spin field} in the sense that it has an infinite
number of physical degrees of freedom \cite{BekSk}.}. Such a new
gauge algebra will be constructed in the present paper and, as we
will see, it differs from gauge algebra for free massless higher
spin field theory in AdS space (see e.g. \cite{BPT} where the BRST
approach was developed for such a theory). Lagrangian formulation is
then realized in the framework of the BRST-BFV construction
on the base of new gauge algebra under consideration. As we will
see, finding such non-minimal terms in the gauge algebra is an
extremely non-trivial problem. Besides, unlike the approaches in
refs. \cite{Met,Zin}, we use the fields on four-dimensional space
with only two-component spinor indices.

The paper is organized as follows. In the section 2 we briefly
describe the Lagrangian construction for free bosonic infinite spin
field in $4D$ Minkowski space in terms of two-component spin-tensor
fields \cite{BKT}. As we will see, the use of such fields essentially
simplifies the BFV construction (see e.g. \cite{BuchKout}). Section
3 is devoted to defining the properties of basic operators which are
used then to formulate the constraints in the AdS space. In section 4
we deduce the new gauge algebra for bosonic infinite spin field
and construct the Lagrangian formulation for this field. The
Lagrangian formulation is consistently derived from unique general
principle without appeal to Fronsdal Lagrangian for massless finite spin
fields and for special limit in higher spin massive theory. In
section 5 we summarize our results.

\section{Bosonic infinite spin field in Minkowski space}\label{sec:mink}
Before moving on to the BRST construction for an infinite spin field in curved space-time,
we will briefly discuss the corresponding construction in flat space.
The covariant description of a real bosonic infinite spin field in Minkowski space within the BRST approach
was presented in \cite{BKT} and then it was generalized for more complicated cases in \cite{BFIK}.
Further we summarize the results obtained there for the subsequent formulation of a bosonic infinite
spin field in $AdS_4$ space.

We describe the bosonic infinite spin field by an infinite collection of spin-tensor fields
(see \cite{BKT,BFIK} for details)
\be
\varphi_{\alpha(s)\dot\alpha(s)}(x):=\varphi_{\alpha_1\ldots \alpha_s\dot\alpha_1\ldots\dot\alpha_s}(x)\,,
\qquad
s=0,1,2,\ldots,
\ee
which are symmetric with respect all undotted and dotted indices:
$\varphi_{\alpha_1\ldots \alpha_s\dot\alpha_1\ldots\dot\alpha_s}=\varphi_{(\alpha_1\ldots \alpha_s)(\dot\alpha_1\ldots\dot\alpha_s)}$, where
$\alpha=(1,2),\, \dot{\alpha}=(\dot{1},\dot{2}).$
These fields are imbedded into Fock space with the help of the creation $c^\alpha$, $\bar{c}_{\dot\alpha}$
and annihilation $a_\alpha$, $\bar{a}^{\dot\alpha}$  operators
acting on vacuum states $|0\rangle$ and $\langle0|$, $\langle0|0\rangle=1$ according to
\footnote{We use the same notation and conventions as in \cite{BKT,BFIK,Ideas}.}
\be\lb{vac}
\langle0|\bar{c}_{\dot\alpha}=\langle0|c^\alpha=0\,,
\qquad
\bar{a}^{\dot\alpha}|0\rangle=a_\alpha|0\rangle=0
\ee
and having the following nonzero commutation relations
\begin{eqnarray}
[\bar{a}^{\dot\beta},\bar{c}_{\dot\alpha}]
=\delta^{\dot\beta}_{\dot\alpha},
&\qquad&
[a_\beta,c^\alpha]=\delta_\beta^\alpha
\,.
\end{eqnarray}
Under the Hermitian conjugation, the creation and annihilation operators transform into each other, and
the dotted and undotted representations also
transform into each other; therefore,
\be\lb{her}
(a_\alpha)^\dagger=\bar{c}_{\dot\alpha}\,,\qquad
(\bar{a}_{\dot\alpha})^\dagger=c_\alpha\,.
\ee
In this case, the conjugation of vacuum states has a standard form
\be
(|0\rangle)^\dagger=\langle0|\,.
\ee

The Fock space of the system under consideration is formed
by ket-vectors
\begin{equation}\label{GFState}
|\varphi\rangle=
\sum_{s=0}^{\infty}|\varphi_{s}\rangle
\,,\qquad
|\varphi_{s}\rangle:=\frac{1}{s!}\,\varphi_{\alpha(s)}{}^{\dot\alpha(s)}(x)\ c^{\alpha(s)}\,\bar{c}_{\dot\alpha(s)}|0\rangle
\end{equation}
and their conjugate bra-vectors
\be\label{GFState-b}
\langle\bar{\varphi}|=
\sum_{s=0}^{\infty}\langle\bar{\varphi}_s|\,,\qquad
\langle\bar{\varphi}_s|:= \frac{1}{s!}\,\langle 0|\,\bar{a}^{\dot\alpha(s)}\,a_{\alpha(s)}\ \bar{\varphi}^{\alpha(s)}{}_{\dot\alpha(s)}(x)\,,
\ee
where we have used the notation
\be
c^{\alpha(s)}:=c^{\alpha_1}\ldots c^{\alpha_s}\,,
\quad
\bar{c}_{\dot\alpha(s)}:=\bar{c}_{\dot\alpha_1}\ldots\bar{c}_{\dot\alpha_s}\,,
\qquad
a_{\alpha(s)}:=a_{\alpha_1}\ldots a_{\alpha_s}\,,
\quad
\bar{a}^{\dot\alpha(s)}:=\bar\alpha^{\dot{\alpha}_1}\ldots\bar\alpha^{\dot{\alpha}_s}\,.
\ee
The states \p{GFState} and \p{GFState-b} automatically satisfy the conditions
\be\lb{N-N}
(N-\bar N) \,|\varphi\rangle=0 \,,
\qquad
\langle\bar{\varphi}|\,(N-\bar N)=0
\,,
\ee
where
\begin{equation}
N:=c^\alpha a_\alpha\,,
\qquad
\bar{N}:=\bar{c}_{\dot\alpha}\bar{a}^{\dot\alpha}\,,
\qquad N^\dagger=\bar N\,. \lb{ex-N}
\end{equation}
Therefore, the use of the Fock space formed by the vectors  \p{GFState} and \p{GFState-b}
gives us the resolution of constraint \p{N-N}.

The remaining constraints are used in the construction are given by the operators
\begin{eqnarray}\lb{op-0-0}
\ell_0 &:=& \partial^2\,,
\\ [6pt] \lb{op-1-0}
\ell &:=& i\,(a\sigma^m\bar{a})\,\partial_m\,,
\\ [6pt] \lb{op-t1-0}
\ell^{+}&:=& i\,(c\sigma^m\bar{c})\,\partial_m,
\end{eqnarray}
where
\be
(a\sigma^m\bar{a})\ := a^\alpha\sigma^m_{\alpha\dot\alpha}\bar{a}^{\dot\alpha} \,,\qquad
(c\sigma^m\bar{c})\ := c^\alpha\sigma^m_{\alpha\dot\alpha}\bar{c}^{\dot\alpha}\,.
\ee
The only non-zero commutator of the operators \p{op-0-0}-\p{op-t1-0} is
\begin{equation}\label{alg(0)}
[\ell^{+},\ell]=K\,\ell_0\,,
\end{equation}
where
\begin{equation}
K:=N +\bar{N}+2\,.
\lb{ex-K}
\end{equation}

Let us now describe the Lagrangian construction within the BRST method (for more details see \cite{BKT}).
Firstly, we construct a BRST charge using the following of operators:
\be
\ell_0\,, \qquad \ell-\bm{\mu}\,, \qquad \ell^{+}-\bm{\mu}
\ee
as the first class constraints satisfying algebra \eqref{alg(0)},
where the real parameter $\bm{\mu}$ indicates a infinite spin representation.
As a result of this standard procedure, the resulting BRST charge has the form
\be\label{Q(0)}
Q \ = \ \eta_0\ell_0 \ + \ \eta^+\left(\ell-\bm{\mu}\right) \ + \ \eta\left(\ell^{+}-\bm{\mu}\right) \ + \
K\eta^+\eta\, \mathcal{P}_0 \,.
\ee
Here we have extended the Fock space by introducing $\eta_0$, $\eta$, $\eta^+$ which are the
fermionic ghost ``coordinates'' and
$\mathcal{P}_0$, $\mathcal{P}^+$, $\mathcal{P}$ which are their canonically conjugated ghost ``momenta'', respectively.
These operators obey the anticommutation relations
\be
\{\eta,\mathcal{P}^+\}
 \ = \ \{\mathcal{P}, \eta^+\}
 \ = \ \{\eta_0,\mathcal{P}_0\}
 \ = \ 1
\label{ghosts}
\ee
and act on the vacuum state as follows
\begin{eqnarray}
&&
\eta|0\rangle \ = \ \mathcal{P}|0\rangle \ = \ \mathcal{P}_0|0\rangle \ = \ 0.
\end{eqnarray}
All the ghost operators possess the standard  ghost numbers,
$gh(\eta_0)= gh(\eta)=gh(\eta^+)=1$ and $gh(\mathcal{P}_0)=gh(\mathcal{P})=gh(\mathcal{P}^+)=-1$, providing the property  $gh(Q)$ = $1$.

The operator \eqref{Q(0)} acts in the extended Fock space of the vectors
\be
|\Phi\rangle \ = \
|\varphi\rangle \ + \ \eta_0\mathcal{P}^+|\varphi_1\rangle \ + \ \eta^+\mathcal{P}^+|\varphi_2\rangle
\label{extened vector}
\ee
having a zero ghost number.
These states obey the BRST equations of motion
\begin{eqnarray}\label{eqQ}
Q\,|\Phi\rangle \ = \ 0\,.
\end{eqnarray}
and are defined up to gauge transformations
\begin{eqnarray}
|\Phi'\rangle \ = \ |\Phi\rangle \ + \ Q\,|\Lambda\rangle \,,
\label{gauge transf}
\end{eqnarray}
where $|\Lambda\rangle$ is the Fock space valued gauge parameter of the form
\begin{eqnarray}\label{gtQ}
|\Lambda\rangle \ = \ \mathcal{P}^+|\lambda\rangle\,,
\label{gauge parameter}
\end{eqnarray}
that has a ghost number equal to $-1$.
The fields $|\varphi_1\rangle$, $|\varphi_2\rangle$ and the gauge parameter  $|\lambda\rangle$
in relations (\ref{extened vector}),
(\ref{gauge parameter}) have a similar decomposition like
$|\varphi\rangle$ in \eqref{GFState}.

The Lagrangian for the bosonic infinite spin field is constructed in the framework of the BRST approach
as follows \cite{BKT}:
\be
{\cal L}
\ = \
\int d\eta_0\; \langle\Phi|\,Q\,|\Phi\rangle
\,.
\label{actionQ}
\ee
The explicit component form of the equations of motion \eqref{eqQ}, gauge transformations \eqref{gtQ} and
component Lagrangian corresponding to the BRST Lagrangian \eqref{actionQ} is given in \cite{BKT}.
One can show that the Lagrangian \eqref{actionQ} is the sum of Lagrangians for massless bosonic fields
plus $\bm{\mu}$-dependent cross terms responsible for the infinite spin field description.

In the case of $\bm{\mu}=0$, the BRST charge (\ref{Q(0)}) becomes one for the infinite collections of massless higher
spin fields and therefore in this case
the Lagrangian \eqref{actionQ} becomes one
of the massless fields of all helicities (see \cite{BuchKout}) where such a construction was described for arbitrary helicity).

Further, we will derive the Lagrangian for the bosonic infinite spin field in AdS space in the framework
of the BRST construction. However, as an intermediate step we will briefly discuss the generalization
of the above constraint operators on curved space.

\setcounter{equation}{0}
\section{Basic operators of constraints in $AdS_4$}\label{sec:ads0}

In order to generalize the above construction on curved space, we first of all need to generalize
operators \p{op-0-0}-\p{op-t1-0} and algebra \eqref{alg(0)}. For this purpose, we introduce the
vielbein $e_{\mu}{}^m(x)$ and spin-connection $\omega_\mu{}^{mn}(x)=-\omega_\mu{}^{nm}(x)$, where
the Greek indices $\mu,\nu$ are the curve space ones and the Latin indices $m,n$  are the tangent
space ones which are lowered and raised by flat metrics $\eta_{mn}$ and $\eta^{mn}$.

A natural generalization of the operators \p{op-1-0}, \p{op-t1-0} looks like
\begin{eqnarray}
\lb{op-1}
l &:=& i\,(a\sigma^m\bar{a})\,e^\mu{}_m D_\mu\,,
\\ [6pt] \lb{op-t1}
l^+&:=& i\,(c\sigma^m\bar{c})\,e^\mu{}_m D_\mu\,.
\end{eqnarray}
Here the  $D_\mu$ are the covariant derivative operators defined by
\be\lb{D-t}
D_\mu\ =\ \partial_\mu \ + \ \frac{1}{2}\ \omega_\mu{}^{mn}\,\mathcal{M}_{mn}\,,
\ee
where\footnote{We use the notation
$\sigma_{mn}=-\frac{1}{4}(\sigma_{m}\bar{\sigma}_{n}-\sigma_{n}\bar{\sigma}_{m})$,
$\tilde{\sigma}_{mn}=-\frac{1}{4}(\tilde{\sigma}_{m}\sigma_{n}-\tilde{\sigma}_{n}\sigma_{m})$.}
\be\lb{M-op}
\mathcal{M}_{mn} \ = \ {M}_{mn}+\bar{M}_{mn}\,, \qquad
{M}_{mn}= c^\alpha \,(\sigma_{mn})_\alpha{}^\beta\,a_\beta \,,\quad
\bar{M}_{mn}=\bar{c}_{\dot{\alpha}}\,(\tilde{\sigma}_{mn})^{\dot{\alpha}}{}_{\dot{\beta}}\,\bar{a}^{\dot{\beta}}
\ee
are the generators of the Lorentz algebra
$$
[\mathcal{M}_{mn},\mathcal{M}_{kl}]=\eta_{ml}\mathcal{M}_{nk}+\eta_{nk}\mathcal{M}_{ml}-\eta_{mk}\mathcal{M}_{nl}-\eta_{nl}\mathcal{M}_{mk}\,.
$$

Further, we consider the torsion-free case where the torsion tensor $T_{\mu\nu}{}^k$ vanishes
and the case of a covariantly constant vielbein.
The last condition allows us to express the Christoffel symbols in a standard way in terms of the vielbein and
spin connection.

It is important that the operators \p{op-1}, \p{op-t1} are Hermitian conjugate to each other
with respect to inner product discussed in Appendix,
$(l)^\dagger=l^+$.

Let us study an algebra of the operators (\ref{op-1}) and \p{op-t1} and
calculate the commutator $[l^+,l]$.
We begin with the commutator
\begin{equation}
\label{D-mu-alg}
\left[ D_\mu ,D_\nu \right] \ = \ \frac{1}{2}\,R_{\mu\nu}{}^{mn}\mathcal{M}_{mn}\,,
\end{equation}
where
\begin{equation}
\label{R-expr}
R_{\mu\nu}{}^{m}{}_{n} \ = \ \partial_\mu\omega_\nu{}^{m}{}_n -\partial_\nu\omega_\mu{}^{m}{}_n
+[\omega_\mu,\omega_\nu]^{m}{}_n
\end{equation}
is the curvature tensor.
In addition, the commutators
\begin{equation}
\label{a-c-alg}
\left[  (c\sigma^m\bar c), (a\sigma^n\bar a)\right]  =  -\, K \eta^{mn}- 2\mathcal{M}^{mn}\,,
\end{equation}
\begin{equation}
\label{M-ac-alg}
\left[ \mathcal{M}_{mn} ,(a\sigma_k\bar a) \right]  =  \eta_{nk}(a\sigma_m\bar a) - \eta_{mk}(a\sigma_n\bar a)\,,\quad
\left[ \mathcal{M}_{mn} ,(c\sigma_k\bar c) \right]  =  \eta_{nk}(c\sigma_m\bar c) - \eta_{mk}(c\sigma_n\bar c)\,,
\end{equation}
lead to the following relations:
\begin{equation}
\label{D-ac-alg}
\left[ D_{m} ,(a\sigma_n\bar a) \right]  =   -\, \omega_{m,nk}\,(a\sigma^k\bar a)\,,\quad
\left[ D_{m} ,(c\sigma_n\bar c) \right]  =   -\, \omega_{m,nk}\,(c\sigma^k\bar c)\,,
\end{equation}
where
\begin{equation}
\label{P-a}
D_m \ :=  \ e^\mu{}_m D_\mu
\end{equation}
and $\omega_m{}^{nk}=e^\mu{}_m\omega_\mu{}^{nk}$, $\omega_{m,nk}=\eta_{nl}\eta_{kp}\omega_m{}^{lp}$.
Note that the commutation relations in Riemannian space for the operators
\begin{equation}
\label{aa-cc-mu}
(a\sigma^\mu\bar a) \ = \ (a\sigma^m\bar a)\,e^\mu{}_m\,,\qquad
(c\sigma^\mu\bar c) \ = \ (c\sigma^m\bar c)\,e^\mu{}_m
\end{equation}
have the form
\begin{equation}
\label{D-ac-alg-mu}
\left[ D_{\mu} ,(a\sigma^\nu\bar a) \right]  =   -\, \Gamma_{\mu\lambda}^\nu\,(a\sigma^\lambda\bar a)\,,\quad
\left[ D_{\mu} ,(c\sigma^\nu\bar c) \right]  =   -\, \Gamma_{\mu\lambda}^\nu\,(c\sigma^\lambda\bar c)\,.
\end{equation}

Using \p{D-mu-alg} and \p{D-ac-alg} (or \p{D-ac-alg-mu}),
we obtain
\be
\left[l^+,l \right] \ = \
K D^2\  - \
\frac{1}{2}\,\Big\{(1+\bar N) M^{mn} + (1+N) \bar M^{mn}\Big\} \, R_{mn}{}^{kl} \Big\{ M_{kl} + \bar M_{kl}\Big\}
\,.
\lb{com-ll-5}
\ee
Here
\be\lb{D2}
D^2\ :=\ \eta^{mn}\left(D_m D_n+ \omega_{m,nk}D^k\right) \ = \
g^{\mu\nu}\left(D_\mu D_\nu-\Gamma_{\mu\nu}^\lambda D_\lambda\right) \ =\
\frac{1}{\sqrt{-g}}\, D_\mu \sqrt{-g}g^{\mu\nu}D_\nu \,,
\ee
where as usual
$g=\det g_{\mu\nu}$.
In fact, the operator $D^2$ is a spin generalization of the Laplace-Beltrami operator.

Further we consider curved spaces with constant curvature.
In this case, the curvature tensor has the form
\be
\lb{cur-2}
R_{mn}{}^{kl}
\ = \
\kappa(\delta_m^k\delta_n^l-\delta_m^l\delta_n^k)
\,,
\ee
with a nonzero constant $\kappa$. In the next section we will show that
unlike finite spin field situation, in the case of infinite spin field, the consistency condition of the generator algebra requires only
negative $\kappa$ that corresponds to the $AdS_4$ space.

Taking into account the identities
for the $\sigma$-matrices, we have
\be\lb{id-s}
M^{mn}M_{mn}= - N(N+2) \,,\qquad \bar M^{mn}\bar M_{mn} =  - \bar N(\bar N+2) \,,\qquad M^{mn}\bar M_{mn}=0\,.
\ee
and
\be\lb{id-ca-M}
(c\sigma_{[m} \bar c)(a\sigma_{n]} \bar a )=\bar N M_{mn} + N \bar M_{mn}\,.
\ee

Inserting \p{cur-2} in \p{com-ll-5} and taking into account \p{id-s} and \p{id-ca-M}, one obtains
\be
\left[l^+,l \right] \ = \
K\;l_0
\,,
\lb{com-ll-6}
\ee
where we have introduced the new operator $l_0$ as follows
\be\lb{op-0}
l_0 := D^2 \ + \ \kappa\left(N\bar{N}+N+\bar{N}\right)\,.
\ee
This operator is a curved space-time generalization of the flat space operator \p{op-0-0}.

One can show that the operator  \p{op-0} forms a closed algebra with operators \p{op-1}, \p{op-t1}.
The remaining commutators of this algebra are
\footnote{
Correct curved space generalization of the relations \p{op-1}, \p{op-t1}, \p{op-0} for the finite spin field generator algebra \p{op-1-0}, \p{op-t1-0}, \p{op-0-0}
exists both for the $dS$ space, when $\kappa>0$ and for the $AdS$ space, when $\kappa>0$.
But, as will emphasize in the next section, the correct generalization of the generators
for infinite spin field is possible only for the $AdS$ space.
}
\be\label{algebra-r}
[l,l_0]=2\kappa\,(K+1)\,l\,,\qquad
[l_0,l^+]=2\kappa\,(K-1)\,l^+\,.
\ee

In the next section, we will discuss the correct curved space generalization of the operators $l$ (\ref{op-1}),
 $l^+$ (\ref{op-t1}), $l_0$ (\ref{op-0}) and their algebra (\ref{com-ll-6}), (\ref{algebra-r})
to infinite (continuous) spin case. Generalization based on 
the closure of the generator algebra which uniquely singles out the case of the $AdS$ space. 
Taking into account this result, we will use a term $AdS$  space in the title of the next section.

\setcounter{equation}{0}
\section{Bosonic infinite spin field in $AdS_4$}\label{sec:ads}
In this section, we are going to construct a generalization of the Lagrangian formulation presented in Section \ref{sec:mink} to obtain the Lagrangian formulation for the infinite spin field in $AdS$ space. The consideration is naturally divided into a generalization of gauge algebra and a generalization of the BRST construction.
\subsection{Generalization of gauge algebra}
Turn attention to that the operators $\ell_0$, $\ell-\bm{\mu}$, $\ell^{+}-\bm{\mu}$ forming the closed
algebra for the infinite spin field in flat space include the parameter $\bm{\mu}$ characterizing the infinite
spin representation.  Therefore, our purpose is to generalize the operators $l_{0}$, $l$, $l^{+}$
so as to obtain
operators involving the parameter $\bm{\mu}$ and forming a closed algebra in the curved space. For this
purpose, we define the new operators  $L_0,\, L,\, L^{+}$
in the following way:
\begin{eqnarray} \lb{op-L0}
\ell_0\quad&\rightarrow&\quad L_0 \ = \ l_0+\Delta_0(K)\,,
\\ [6pt] \lb{op-L1}
\ell\quad&\rightarrow&\quad L \ = \ l-\Delta(K)\,,
\\ [6pt] \lb{op-L1t}
\ell^{+}\quad&\rightarrow&\quad L^+ \ = \ l^+-\Delta(K)
\,.
\end{eqnarray}
Here the operators $l_{0}$, $l$, $l^{+}$ are defined in section ~\ref{sec:ads0}. As for additional
operators, one requires that $\Delta_{0}(K)$ vanishes at $\kappa=0$ and
\be\lb{DK}
\Delta(K)=\bm{\mu}+\ldots \,,
\ee
where the dots mean the terms that disappear at $\kappa=0$.
The main requirement to the new operators is that their algebra is closed. Let us move on
to constructing such an algebra.

Let us begin with the commutator $[L^+,L]$ which equals
\begin{eqnarray}
[L^+,L]
&=&
Kl_0
+\Bigl[\Delta(K+2)-\Delta(K)\Bigr]\;l
+\Bigl[\Delta(K)-\Delta(K-2)\Bigr]\;l^+
=
\nonumber
\\
&=&
KL_0
+\Bigl[\Delta(K+2)-\Delta(K)\Bigr]\;L
+\Bigl[\Delta(K)-\Delta(K-2)\Bigr]\;L^+ \,,
\end{eqnarray}
where the operator $K$ is defined in section ~\ref{sec:mink}. Here we have used the relations
$$
lK=(K+2)l\,,\qquad Kl^+=l^+(K+2).
$$
Besides, we
require here that the operators $\Delta_0(K)$ and $\Delta(K)$ satisfy the relation
\be
K\Delta_0(K)
=
\bigl[\Delta(K+2)-\Delta(K-2)\bigr]\Delta(K)\,.
\label{Delta0}
\ee

Now one considers the commutator $[L,L_0]$. After some transformations we get
\begin{eqnarray}
[L,L_0]
&=&
2\kappa\,(K+1)\,L
+\Bigl[\Delta_0(K+2)-\Delta_0(K)\Bigr]\;L
-
\nonumber
\\
&&{}
+2\kappa\,(K+1)\,\Delta(K)
+\Bigl[\Delta_0(K+2)-\Delta_0(K)\Bigr]\;\Delta(K)
\,. \lb{com10}
\end{eqnarray}
For the algebra to be closed, the second row on the right hand side in \p{com10} must vanish:
\be
2\kappa\,(K+1)\,\Delta(K)
+\Bigl[\Delta_0(K+2)-\Delta_0(K)\Bigr]\;\Delta(K)
=0
\,.
\ee
However, due to \p{DK}, $\Delta(K)\neq 0$.
Therefore, we need to require that the following equation should be fulfilled
\be\lb{cond1}
\Delta_0(K+2)-\Delta_0(K)
=
-2\kappa\,(K+1)
\,.
\ee
Solution to this equation is
\begin{eqnarray}
\Delta_0(K)=A+2\kappa-\frac{1}{2}\,\kappa K^2
\,,
\label{D0K}
\end{eqnarray}
where $A$ is an arbitrary constant. Note that if the condition \p{cond1} is satisfied, then the
commutator \p{com10} is equal to zero, $[L,L_0]=0$.

The next step is finding the $\Delta(K)$. For this aim we introduce the operator
\be\lb{fK}
f(K)\ =\ \Delta(K+2)\Delta(K)\,.
\ee
Using relations \eqref{Delta0} and \eqref{D0K}, one can show that the operator $f(K)$ satisfies the equation
\be
f(K)
\ - \
f(K-2)
\ = \
AK \ - \ \frac{1}{2}\,\kappa\,K(K+2)(K-2)
\,.
\ee
It is easy to check that the solution to this equation has the form:
\be
f(K)
\ = \
B
 \ + \ \frac{1}{4}\, AK(K+2)
 \ - \ \frac{1}{16}\,\kappa\,K(K+2)(K-2)(K+4)
\,,
\label{fK-eq}
\ee
where $B$ is another arbitrary constant.
Relation \p{fK-eq} leads to
\be
\Delta(K+2)\Delta(K)
\ = \
B
 \ + \ \frac{1}{4}\, AK(K+2)
 \ - \ \frac{1}{16}\,\kappa\,K(K+2)(K-2)(K+4)
\,.
\label{15}
\ee

We will look for the function $\Delta(K)$ on $K$ in the form of a power series in $K$.
Then, after some transformations, one gets the following solution to equation \p{15}:
\be\lb{Delta}
\Delta(K)\ = \ a + b K^2\,,
\ee
where the constants $a$, $b$ and the constants  $A$, $B$  are limited by the conditions
\begin{eqnarray}
b^2&=& -\,\kappa/16\,, \lb{c-b}
\\ [6pt]
A&=& 2\left(4ab-\kappa \right)\,, \lb{c-A}
\\ [6pt]
B&=& 4ab+a^2\,. \lb{c-B}
\end{eqnarray}
Thus, only one additional condition can be imposed to fix the four constants $a$, $b$, $A$, $B$. Turn
attention to that the condition \eqref{c-b} shows that the constant $\kappa$ must be negative:
\be\lb{kappa}
\kappa\ < \ 0\,.
\ee
As a result we conclude that the constant curvature space under consideration must actually be the
AdS space.\footnote{Note that the operators $L$ \p{op-L1} and $L^+$ \p{op-L1t},
defining the constraints in our construction, are conjugate to each other with respect
to the Hermitian conjugation \p{her}.
This condition is fundamental in the construction of the Hermitian BRST operator and, hence, for the real Lagrangian.
Within the framework of the construction of field theory under consideration,
we can make the only relaxation in definition of the operators $L$ and $L^+$:
we can take the constraints $L = l-\Delta(K)$, $L^+  = l^+-\bar{\,\Delta}(K)$
instead of the constraints \p{op-L1} and \p{op-L1t}.
As solution to $\Delta(K)$, we will again obtain the expression \p{Delta}
where the coefficients $a$ and $b$ are now complex numbers.
In this case, instead of the condition \p{c-b}, the condition $b\bar b= -\,\kappa/16$ must take place.
But this condition is met only for negative values of the constant $\kappa$.
That is, the approach under consideration is consistent only for the AdS case \p{kappa}.
Our result represents another independent proof of the statements in \cite{Met,Met2,Zin} that the infinite spin field theory, unlike 
the finite spin field theory, is consistently formulated only in the AdS space.}

Although in the general case $b=\pm |\kappa|^{1/2}/4$, for definiteness we take
\be
b \ = \ |\kappa|^{1/2}/4\,.
\ee

Taking into account the limit \p{DK}, we can put
\be
a \ = \ \bm{\mu}
\ee
as the fourth condition to determine the constants $a$, $b$, $A$, $B$.
Then
\be
A=-2\kappa + 2\,\bm{\mu}\,|\kappa|^{1/2}
\ee
and
\begin{eqnarray}
\Delta_0(K)&=&2\,\bm{\mu}\,|\kappa|^{1/2} \, - \,  \frac{1}{2}\,\kappa K^2
\,, \\ [6pt]
\Delta(K)&=&\bm{\mu} \,  + \,  \frac{1}{4}\,|\kappa|^{1/2}K^2
\,.
\end{eqnarray}

As a result, we find all the operators \p{op-L0}-\p{op-L1t}:
\begin{eqnarray}
L_0&=&l_0 \, + \, 2\bm{\mu}\,|\kappa|^{1/2} \, - \, \frac{1}{2}\,\kappa K^2
\,,\lb{op-L0-f}
\\ [6pt]
L&=&l \, - \, \bm{\mu}  \, - \, \frac{1}{4}\,|\kappa|^{1/2}K^2\,,
\lb{op-L1-f}
\\ [6pt]
L^+&=&l^+ \, - \, \bm{\mu} \,  - \, \frac{1}{4}\,|\kappa|^{1/2}K^2 \,.
\lb{op-L1t-f}
\end{eqnarray}
The corresponding commutator algebra for these operators is written in the form
\begin{eqnarray}
\lb{al1}
[L,L_0]&=&0\,,
\qquad
[L^+,L_0] \ = \ 0\,,
\\ [7pt]
[L^+,L]
&=&
KL_0
+|\kappa|^{1/2}(K+1)L
+|\kappa|^{1/2}(K-1)L^+ \,.
\lb{al2}
\end{eqnarray}

To show that the right hand side of relation ~(\ref{al2}) is Hermitian, one uses the relations
\be
LK=(K+2)L+2\Delta(K)\,,\qquad KL^+=L^+(K+2)+2\Delta(K)\,.
\ee
After that, the right-hand side of the commutator $[L^+,L]$  is rewritten in explicitly Hermitian form:
\begin{eqnarray}
[L^+,L]&=&
KL_0
+\frac{1}{2}\,|\kappa|^{1/2}(K+1)L
+\frac{1}{2}\,|\kappa|^{1/2}(K-1)L^+
+
\nonumber
\\[6pt]
&&\hspace{15ex}{}
+\frac{1}{2}\,|\kappa|^{1/2}L^+(K+1)
+\frac{1}{2}\,|\kappa|^{1/2}L(K-1)\,.
\label{L1+L1erm}
\end{eqnarray}
Relations \p{al1} and ~(\ref{al2}) are the final form of gauge algebra for infinite spin field
in $AdS_4$. It is easy to see that at $\kappa=0$ this algebra coincides with gauge algebra for
the infinite spin field in flat space. Further, we will apply the above algebra for deriving the
Lagrangian formulation for the infinite spin field in the AdS space.

\subsection{Lagrangian construction}

The Lagrangian description of the infinite spin field in the $AdS_4$
space is developed following a generic procedure similarly to how it was
done in section\,2 for the field of infinite spin in flat space.
The Lagrangian formulation is constructed in the framework of the BRST
approach in the Fock space of the vectors \p{GFState} and
\p{GFState-b} using the constraints \p{op-L0-f}, \p{op-L1-f},
\p{op-L1t-f} that generate the commutator algebra \p{al1}, \p{al2}.

The central object in the BRST approach to the Lagrangian construction for
higher spin theories is the BRST charge $Q$ that should satisfy two basic
requirements, Hermitianity and nilpotency:
 \begin{eqnarray}
Q^+=Q\,,
 \lb{Q1}
 \\
 [6pt]
 Q^2=0\,.
 \lb{Q2}
 \end{eqnarray}
Hermitianity provides the reality of the Lagrangian, and nilpotency
provides gauge invariance.

Consider a construction of the BRST charge satisfying the conditions
\p{Q1} and \p{Q2}. We begin with the standard prescription for
constructing the BRST charge using the constraints \p{op-L0-f},
\p{op-L1-f}, \p{op-L1t-f}
\be
 Q_{naive} \ = \ \eta_0L_0+\eta^+L+\eta L^+
+K\eta^+\eta\mathcal{P}_0
+|\kappa|^{1/2}(K+1)\;\eta^+\eta\mathcal{P}
+|\kappa|^{1/2}(K-1)\;\eta^+\eta\mathcal{P}^+ \,. \label{Q0}
\ee
However, the standard expression \p{Q0} is too formal in the sense
that it does not take into account the operator structure of
commutator algebra \p{al1}, \p{al2}. Therefore, in general, the BRST
charge \p{Q0} does not have to be either Hermitian or nilpotent. In
other words, the true BRST charge as an operator acting in Fock space
and satisfying the conditions \p{Q1} and \p{Q2} needs additional
definitions. We will call expression \p{Q0} a naive BRST
charge. Our main aim is to propose such additional definitions for
the operator \p{Q0} to construct on its basis a true BRST charge
satisfying the basic conditions \p{Q1} and \p{Q2}.

First of all we observe  that the operator $Q_{naive}$ \p{Q0} in our case is
automatically nilpotent
\be
Q_{naive}^2 \ = \ 0\,,
\ee
but it is not Hermitian: the Hermitian conjugate operator
\be
Q_{naive}^{\dagger} \ = \ \eta_0L_0+\eta^+L+\eta L^+
+K\eta^+\eta\mathcal{P}_0
+|\kappa|^{1/2}(K+1)\;\mathcal{P}^+\eta^+\eta
+|\kappa|^{1/2}(K-1)\;\mathcal{P}\eta^+\eta \label{Q00}
\ee
is not equal to the operator \p{Q0}, $Q_{naive}^{\dagger} \neq
Q_{naive} $ due to the presence in \p{Q0} and \p{Q00} of the terms
containing the products $\eta^+\mathcal{P}$ and $\eta\mathcal{P}^+$
with non-commuting pairs of operators $\eta^+$, $\mathcal{P}$ and
$\eta$, $\mathcal{P}^+$.

To provide Hermitianity, we will attempt to define the true BRST
charge $Q$ in the form
\be\lb{Q+Qb}
 \frac12\left(Q_{naive}+Q_{naive}^{\dagger}\right) .
\ee
This operator is automatically Hermitian. However, it is not
nilpotent. We should introduce an additional definition.

Now let us pay attention to the fact that the operator \p{Q+Qb}
needs to specify the ordering  prescriptions for the non-commuting
creation and annihilation operators acting in the Fock space. To
achieve nilpotency, we will use in expression \p{Q+Qb}
the symmetric ordering prescription for product of the creation and annihilation operators
\be\lb{ord-b}
c^\alpha a_\beta\quad\rightarrow\quad :c^\alpha a_\beta\!: \
=\frac12\left(c^\alpha a_\beta+a_\beta c^\alpha \right), \qquad\quad
\bar{c}_{\dot\alpha}\bar{a}^{\dot\beta}\quad\rightarrow\quad
:\bar{c}_{\dot\alpha}\bar{a}^{\dot\beta}\!: \ =\frac12
\left(\bar{c}_{\dot\alpha}\bar{a}^{\dot\beta}+\bar{a}^{\dot\beta}\bar{c}_{\dot\alpha}\right).
\ee
This ordering is formally equivalent to replacements (see \p{ex-N})
\be
N\quad\rightarrow\quad :N\!: \ = N+1\,,\qquad\qquad \bar N
\quad\rightarrow\quad :{\bar N}\!: \ =\bar N-1\,
\ee
Here and bellow the denotation $:..:$ means the ordering prescription \p{ord-b}. As a result, we obtain the following ordering prescription (see \p{op-1})
\be
L_0\quad\rightarrow\quad :L_0\!: \ =L_0-\kappa -N + \bar N\,.
\ee
However, when one constructs the Lagrangian on the basis of the BRST
charge according to definition \p{actionQ}, we use the vectors
\p{GFState} and \p{GFState-b} for which the equality $N =
\bar N$ is automatically fulfilled (see \p{N-N}). Therefore, finally
the ordering effect in the BRST charge means a simple replacement
\be
\lb{ord-L}
L_0\quad\rightarrow\quad :L_0\!: \ =L_0-\kappa \,.
\ee

Thus, the true BRST charge is defined as the ordering of the
operator \p{Q+Qb}:
\be
\lb{Q+Qb-def}
 Q \ = \
:\frac12(Q_{naive}+Q_{naive}^{\dagger}): \,.
\ee
Note
that in \p{Q+Qb-def} the ordering $:...:$ is taken with respect
to the bosonic creation and annihilation operators, as in \p{ord-b} and
\p{ord-L}, ordering of ghost operators is not used.

As a result, we come to the following definition of the true BRST
charge for the field under consideration in the $AdS_4$ space:
\begin{eqnarray}
Q&=&
\eta_0(L_0-\kappa)\ + \ \eta^+L \ + \ \eta L^+
+K\eta^+\eta\mathcal{P}_0
+
\nonumber
\\[0.4em]
&&\qquad{}
+\frac{1}{2}\,|\kappa|^{1/2}(K+1)\;\eta^+\eta\mathcal{P}
 \ + \ \frac{1}{2}\,|\kappa|^{1/2}(K-1)\;\eta^+\eta\mathcal{P}^+
+
\nonumber
\\[0.4em]
&&\qquad{}
+\frac{1}{2}\,|\kappa|^{1/2}(K+1)\;\mathcal{P}^+\eta^+\eta
 \ + \ \frac{1}{2}\,|\kappa|^{1/2}(K-1)\;\mathcal{P}\eta^+\eta
\,.
\label{Qfin}
\end{eqnarray}
The BRST charge defined this way is automatically nilpotent and
Hermitian:
\be
 Q^2=0\,, \qquad Q^+=Q\,.
 \ee
Relations (\ref{Q+Qb}) and (\ref{ord-b}) should be considered as part of the definition of the true
BRST charge on the basis of naive expression (\ref{Q0}).

Since the operator (\ref{Qfin}) satisfies the basic properties \p{Q1}
and \p{Q2} of the true BRST charge, we can use it to construct the
Lagrangian
\begin{eqnarray}
\mathcal{L}
&=&
\langle\bar\varphi|\,\left\{
(L_0-\kappa)|\varphi\rangle-\left[L^++\frac{1}{2}\,|\kappa|^{1/2}(K-1)\right]|\varphi_1\rangle
\right\}
\nonumber
\\ [6pt]
&&{}
-\langle\bar\varphi_{1}|\,\left\{
\left[L+\frac{1}{2}\,|\kappa|^{1/2}(K-1)\right]|\varphi\rangle
-\left[L^+-\frac{1}{2}\,|\kappa|^{1/2}(K+1)\right]|\varphi_2\rangle
+K|\varphi_1\rangle
\right\}
\nonumber
\\ [6pt]
&&{}
-\langle\bar\varphi_{2}|\,\left\{
(L_0-\kappa)|\varphi_2\rangle
-\left[L-\frac{1}{2}\,|\kappa|^{1/2}(K+1)\right]|\varphi_1\rangle
\right\}\,.
\label{action-mr}
\end{eqnarray}
Due to nilpotency of the BRST charge (\ref{Qfin}), the Lagrangian
\p{action-mr} is automatically invariant under the gauge transformations
\begin{eqnarray}
\delta|\varphi\rangle &=& \Bigl[L^++\frac{1}{2}\,|\kappa|^{1/2}(K-1)\Bigr]|\lambda\rangle\,,
\label{gt-1}
\\ [6pt]
\delta|\varphi_1\rangle &=& (L_0-\kappa)|\lambda\rangle\,,
\\ [6pt]
\delta|\varphi_2\rangle &=& \Bigl[L-\frac{1}{2}\,|\kappa|^{1/2}(K+1)\Bigr]|\lambda\rangle
\,.
\label{gt-2}
\end{eqnarray}
Relations (\ref{Qfin}), \p{action-mr}, (\ref{gt-1})-(\ref{gt-2}) are our final results.

Note that Lagrangian \eqref{action-mr} and gauge transformations \eqref{gt-1}-\eqref{gt-2} at
$\kappa=0$ turn to Lagrangian and gauge transformations for the bosonic infinite spin in Minkowski space.

\setcounter{equation}{0}

\section{Summary and outlook}\label{ack}
In this paper, we have developed the Lagrangian description of the bosonic infinite spin field
in $AdS_4$ space. The description is based on the BRST construction, where the Lagrangian is formulated
in terms of the operator BRST charge acting in the Fock space corresponding to the creation and annihilation
operators with two-component spinor indices. It its turn, the BRST charge is built with the help of operator
constraints forming first class algebra.

We have generalized the operator constraints, which correspond to the infinite spin field theory in
flat space, to curved space-time and obtained a system of operators \p{op-L0-f}-\p{op-L1t-f} that
determine the operator constraints of the infinite spin fields in $AdS$ space. In flat limit,
these operators coincide with the flat space constraints \p{op-0-0}-\p{op-t1-0}.
The operators \p{op-L0-f}-\p{op-L1t-f} form a closed algebra in terms of commutators which can be treated
as a gauge algebra for infinite spin field theory in $AdS_4$ space. Using these constraints and special
ordering prescription for non-commuting constraints and the creation and annihilation operators we have
constructed the Hermitian and nilpotent BRST charge \p{Qfin} which allows us derive the
Lagrangian {(\ref{action-mr})} and gauge transformations {(\ref{gt-1})-(\ref{gt-2})} for the theory under consideration. The obtained Lagrangian and gauge transformations are formulated in terms of spin-tensor
fields and have the triplet structure.

The results obtained can be considered as a basis for further
development in various directions. First, formulation of fermionic
infinite field theory, second, derivation of interaction vertices
for bosonic and fermionic infinite spin fields and third,
construction of supersymmetric infinite spin filed theory in $AdS_4$
space. In particular, the results obtained open a
possibility to find the manifestly covariant cubic interaction
vertex for infinite spin fields on AdS space among themselves and
for their interactions with massless finite spin fields. Although some
aspects of cubic interactions with infinite spin fields are
discussed in the literature (see e.g., \cite{Met1} and the
references therein), a problem of their interactions in the AdS space is
completely open. We note that in our approach, this problem can be
studied from general view point as a deformation problem in the
framework of the BFV construction.

The triplet structure of the Lagrangian and gauge
transformations, found in the paper, is very convenient for various
generalizations. In particular, such a structure is well adjusted
with supersymmetry transformations \cite{BFIK}. Taking this into
account, we plan to develop a superfield Lagrangian formulation in
$AdS_4$ in a similar way to what was done in \cite{BFIK} for the
case of Minkowski space.

\renewcommand\theequation{A.\arabic{equation}} \setcounter{equation}0
\section*{Appendix \\ Hermitian properties of operators in curved space}

In this Appendix we discuss the aspects of Hermitian conjugation for the operators acting on
the Fock space vectors taking values in curved space-time. The inner product of two such vectors
$|\Psi\rangle$ and $|\Phi\rangle$ is defined as follows
\be\lb{sc-pr}
(\Psi\,|\,\Phi)= \int d^4x \sqrt{-g}\ \langle\Psi(x)\,|\,\Phi(x)\rangle\,,
\ee
where $\langle\Psi(x)|\Phi(x)\rangle$ is the inner product in the Fock space formed by spinor oscillators
defined in \p{vac}.

From the definition \p{sc-pr} we obtain
\be\lb{sc-pr-par}
\int d^4x \sqrt{-g}\ \langle\Psi(x)\,|\,\partial_\mu\Phi(x)\rangle \ = \
-\int d^4x \sqrt{-g}\ \langle(\partial_\mu+\Gamma^\lambda{}_{\mu\lambda})\Psi(x)\,|\,\Phi(x)\rangle \,,
\ee
where $\Gamma^\lambda{}_{\mu\lambda}=\partial_\mu g/(2g)$.
This yields the Hermitian property:
\be\lb{Her-D}
(i D_\mu)^\dagger \ = \
i D_\mu+i\Gamma^\lambda{}_{\mu\lambda} \,.
\ee

Then, the operators \p{op-1}, \p{op-t1} satisfy
\be
\lb{op-1-ap}
(l)^\dagger = \Big(i D_\mu\Big)^\dagger\,\Big((a\sigma^m\bar{a})\,e^\mu_m\Big)^\dagger
= i\,(D_\mu+\Gamma^\lambda{}_{\mu\lambda})(c\sigma^m\bar{c})\,e^\mu_m
\ee
since
\be
\lb{her-aa-cc}
\Big((a\sigma^m\bar{a})\,e^\mu_m \Big)^\dagger=
(c\sigma^m\bar{c})\,e^\mu_m\,.
\ee
Taking into account \p{D-ac-alg-mu} in \p{op-1-ap} one gets
\be
\lb{ld-1}
(l)^\dagger \ = \ l^+\,,
\ee
i.e., the operators $l$ and $l^+$ are Hermitian conjugate to each other.

Using \p{Her-D} we see that the operator \p{D2} is Hermitian:
\be
\lb{D2-her}
(D^2)^\dagger \ = \ D^2\,.
\ee

\bigskip
\section*{Declaration of competing interest}
The authors declare that they have no financial interests/personal relationships that could be considered as potential competing interests.

\bigskip
\section*{Data Availability Statement}
No Data are associated with the manuscript.

\end{document}